\documentclass[USenglish,twocolumn]{article}

\usepackage[utf8]{inputenc}
\usepackage[big]{dgruyter}

\begin{document}

  \articletype{Review Article{\hfill}Open Access}

  \author*[1]{Luka \v C. Popovi\'c}

  \affil[1]{Astronomical Observatory, Volgina 7, 11160 Belgrade, Serbia, E-mail: lpopovic@aob.rs}

  \title{\huge Broad spectral lines in AGNs and supermassive black hole mass measurements}

  \runningtitle{Spectral lines and SMBHs}


  \begin{abstract}
{ The mass measurement of supermassive black holes (SMBHs) is a very complex task. Between
several methods for SMBH mass measurements, some of them use the spectral lines, which indicate the
motion of the emitting/absorbing material around an SMBH. Mostly, there is an assumption of virialization of
line emitting gas in the region which is close to the central SMBH.  In this paper we will give
an overview of methods for the SMBH mass measurements using broad emission spectral lines observed in Type 1 AGNs. First we give the basic idea to use the parameters of broad lines to SMBH mass measurements. After that we give an overview of broad lines from X-ray (Fe k$\alpha$) to the IR (Pashen and Brecket lines) which have been used for SMBH mass estimates. Additionally, we describe and discuss a new method for SMBH mass measurements using the polarization in the broad lines emitted from Type 1 AGNs.
}
\end{abstract}

  \journalname{Open Astronomy}
\DOI{DOI}
  \startpage{1}
  \received{August 10, 2019}
  \revised{December 20, 2019}
  \accepted{?????}

  \journalyear{2020}
  \journalvolume{1}

\maketitle

\section{Introduction}
It is generally accepted that super-massive black holes (SMBHs) are located in  the center of
  massive galaxies. This has been confirmed by recent observations of SMBH shadow in the center of active galactic nuclei (AGN) of M87 galaxy [1]. SMBHs have a significant influence on the structure of the host galaxy, that has been observed as some relationships between the central SMBH mass and their host galaxy structures (see e.g. [2-5],etc.)

The kinematics and physics of matter that surrounds central SMBHs have been strongly affected by their gravitational fields    [6]. Moreover, an SMBH has  in general influence on the  host galaxy formation and evolution [7]. Therefore the investigation of the SMBHs is a very important task in the astrophysics today.

Generally, a black hole  is defined by its mass, spin and electricity, where the mass of  SMBHs seems to be  the most important parameter. Consequently the mass measurements of SMBHs on different cosmic scales and their connections with the structure and evolution of the host galaxies can give more information about  the evolution of our Universe. This implies that estimates (or measurements) of SMBH masses  are very important for cosmological investigation.

However, the measurement of SMBH masses in the center of galaxies represents a very complex task (see e.g. [8]), and there are several methods to perform these measurements. Mostly we divided them into  direct and indirect methods. The direct methods use the dynamics of stars and emitting/absorbing gas which are accelerated by an SMBH itself. On the other side, the indirect methods use some correlations between some parameters of galaxies and their black hole masses (see e.g. [2]). 
In both cases, spectral lines can be used in the SMBH mass measurement.

In the direct methods, we use a fact that stars and gas emit/absorb lines, and that an SMBH has influence on the motion of the stars/gas. Consequently gravitational influence can be observed in the line profiles (as a line shift, or/and in the broadening effects of the line Full Width at Half Maximum -- FWHM). 

If one observes a group of stars around an SMBH, which are emitting an absorption line, the influence of SMBH can be seen in the line dispersion. Therefore,  the relation between stellar dispersion in host bulge stars and SMBH masses  is widely used to estimate SMBH masses (so called M$_\textrm{SMBH}$ -- $\sigma_*$ relationship, see [2, 9-12], etc.).

In the case of AGNs, one can observe emitting gas around an SMBH, that is probably virialized. The emission lines emitted from this gas can be redshifted due to the gravitational redshift, and/or FWHM could reflect the gravitational bounded motion, which depends from the SMBH mass [8,13,14]. These facts are often used to estimate the SMBH masses. Additionally, it has been found, recently, that polarization in broad lines of Type 1 AGNs can be used for SMBH mass measurements (see [15-17]), and that this method gives good agreement in comparison with reverberation and M$_\textrm{SMBH}$ -- $\sigma_*$ methods [17].

In this review, we will discuss the influence of the SMBH mass  to the spectral line shapes emitted from the gas around SMBHs in the case of Type 1 AGNs. The aim of this paper is to give an overview of the  possibilities to use emission lines originated from different regions of an AGN (from the accretion disc to the outside regions near to the inner part of dusty torus) which are emitted in different spectral bands (from the X-ray to the infrared) for SMBH mass measurements. To do that we compile the relations for SMBH mass measurements using broad lines (from different spectral bands) taken form literature. Additionally, we try to give a critical discussion about possibility to use specific broad line for the SMBH mass determination. The paper is organized as following:
In \S 2 we give short overview of the emission lines which can be observed in Type 1 AGNs, \S 3 describes some effects that we expect to see in the spectral line shapes. In \S 4 we give overview of relations between different line parameters in different spectral bands and black hole masses which can be used for SMBH mass estimates. Finally, in \S 5 we give our conclusions with a small critical discussion about accuracies of presented methods.

\section{Spectral lines  in the SMBH vicinity}

An SMBH is mostly located in the center of galaxies, therefore, one expect to have stars and gas around the SMBH. From spectral observations of  active galactic nuclei, it is known that physics and kinematics of emitting gas around black hole could be quite different. 

Very close to the SMBH one expect to have very hot gas, that is in the form of an accretion disc, i.e. gas is accreting in the SMBH. This, hot gas, can produce a very intensive X-ray radiation and very often in  spectra of Type 1 AGNs the  X-ray lines can be detected. One of the most intensive line  in the AGN X-ray spectral band is Fe K$\alpha$ line, with rest energy of $E_0=6.4$ keV 
(see e.g. [18-23], etc.). Since the emission of the Fe K$\alpha$ line can be from the last stable orbit, then the line shape is strongly influenced by the SMBH gravitation and some SMBH parameters are reflecting the Fe K$\alpha$ line shape (see [21] in more details).

An AGN accretion disk is surrounded by a region which emits broad lines, so called the broad line region (BLR), that can be very complex (see e.g. [24]) and different broad lines can come from different emitting regions. For some emitting regions, the physical conditions are such that
high-ionization lines (HILs), for which higher energy is needed to
ionize an atom, can be formed.
As an example C IV $\lambda$1550\AA\ is typical HIL with ionization around 54 eV (see [24]).

On the other side, there are low-ionization lines (LILs), as e.g. Hydrogen Balmer lines and Mg II $\lambda$ 2800 \AA\ which are probably originating in different parts of the BLR [14]. It is expected that the inner part of BLR is emitting HILs (e.g. C IV and C III]). The shape of HILs often indicates some kind of outflows in the BLR, therefore, the geometry of LILs and HILs may be different. Also, there is an emission of Hydrogen Pashen and Brecket series that can be observed in the infrared spectral band. The lines are probably coming from an outer part of the BLR, that can be detected also in Type 2 AGNs (AGNs without broad lines in the optical spectral band), and can be used for measurements of SMBH masses in Type 2 AGNs.

Here we  discuss possibility to measure SMBH masses  using  Fe K$\alpha$, UV, optical and infrared broad lines which are observed  in AGN spectra. 
The list of lines considered to be affected by the central SMBH, and consequently used for the black hole estimate is present in Table 1.

\begin{table*}[h]
\begin{center}
\caption{The list of AGN broad lines which have been used for SMBH mass measurements.}
\label{line-list}
\begin{tabular}{cccccc}
\hline
Emission  &Rest wavelength   &Spectral range & Ref.  \\
       &             &       &   \\
\hline
Fe K$\alpha$& 6.4 keV &   X-ray  & 66-70 \\
Ly$\alpha$& 121.5 nm  &  UV  &    71-76  \\
 C IV & 155 nm  &  UV  & 54,       56, 78-84 \\
CIII] & 190.9 nm &  UV  &      76,84,89,90\\
Fe III UV bump& 204-211 nm  &   UV      & 61,62 \\
Mg II &280 nm & UV &   56,62,76,84,92-97  \\
H$\beta$ and  H$\alpha$& 486.1 nm, 656.3 nm &   optical  & 103-116      \\
Pa$\alpha,\beta$ Br$\alpha,\beta$& 1282-4051 nm &  infrared & 117-121     \\
\hline
\end{tabular}
\end{center}
\end{table*}

\section{Expected influence of SMBH to the spectral line shapes}

An SMBH can be  surrounded by emitting gas, that is accreting into the SMBH in a form of the  accretion disc. The disc emits photons in a wide 
energy range, from X-ray to the far infrared emission. The X-ray and UV photons (with high energies) can photoionize the gas in the BLR that is farther from the central SMBH, but it is still driven by SMBH gravitational force. The photoinized gas emits lines which are affected by this motion and their width and shift can be used for the SMBH mass determination. 

As it is shown in Fig. 1, the photoionized clouds have different velocities, clouds closer to the SMBH (green) should have higher velocities and contribute to the line wings (see Fig. 1 upper left panel), while the clouds farther from the central SMBH are contributing to the central part of a broad line.

\begin{figure*}
\includegraphics[width=0.75\textwidth]{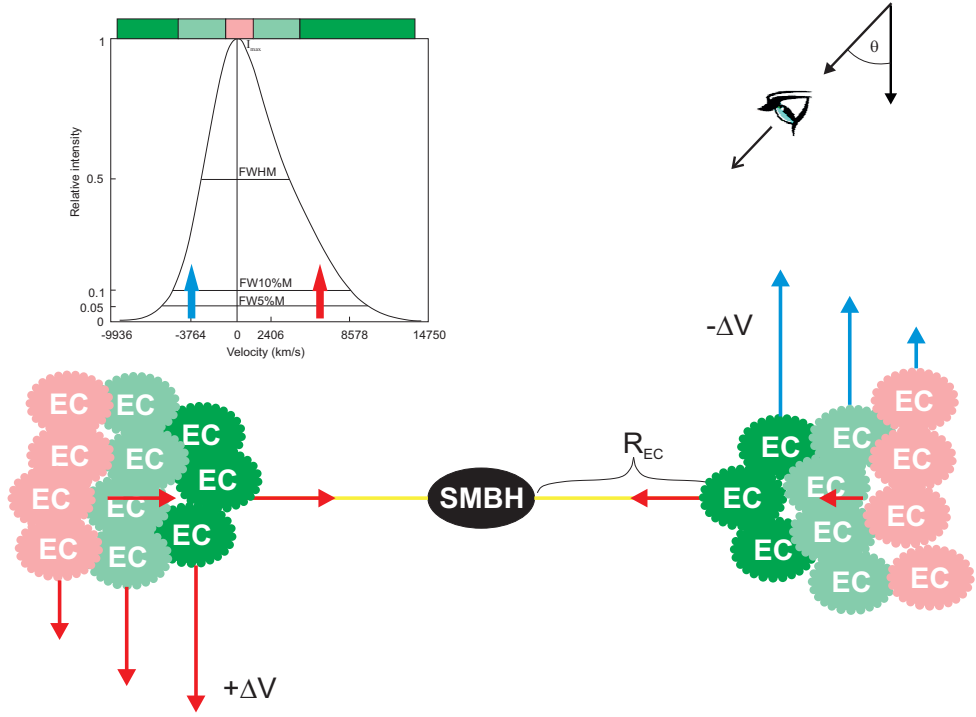}
\caption{Scheme of the  emitting gas motion  in the SMBH vicinity (below) and expected  broad line profile (up) with contributions of different emitting clouds (with rotation-like motion, red arrows denote receding velocities, and blue approaching velocities to the observer). The upper panel shows the line profiles with a red asymmetry, that is assumed to be due to gravitational redshift (red horizontal arrows oriented to the central SMBH).} \label{fig1}
\end{figure*}

On the other side, there is a rotational-like motion of clouds, and  clouds  which are approaching to the observer contribute to the blue line part, and the ones which are receding contribute to the red line part. Then different  random gas velocities in combination with the rotational motion (that is  a function of the distance from the central SMBH,  see Fig. \ref{fig1}) lead to the line broadening. The FWHM reflects the motions of gas which has the rotational component, but also the component of randomly motion of clouds. Both of the velocity components depend on the gravitation field of central SMBH. Therefore, in the case of virialized gas motion the FWHM should be connected with the SMBH mass. Additionally, in the case that the emitting region is close to the SMBH, the line red shift and/or line profile asymmetry  (as it is shown in Fig. \ref{fig1}) caused by the gravitational redshift could be observed (see [25]).

\subsection{Mass of SMBH and line width}

First idea about possibility to use broad spectral lines for SMBH mass measurements in the case of AGN was established in late 1070s, i.e. around ten years after quasar discovery (see [26,27] and critical discussion  in [28]).

Assumption that in the SMBH vicinity the emission gas is virialized  implies that the line FWHM  is reflecting the rotation velocity ($v_\textrm{\textrm{gas}}$) in the BLR due to gravitational motion around the SMBH with mass $M_\textrm{BH}$. The gas velocity is connected with the SMBH mass  as:
$$ v^2_\textrm{gas} \approx {GM_\textrm{BH}\over R},\eqno(1)$$
where $R$ is distance of the emitting region that  emits a broad line, and $G$ is gravitational constant.

As it can be seen in Fig. \ref{fig1} the line-of-sight (LOS) of an observer concerning the rotation velocity can have different angles, i.e. different inclination ($\theta$), then 
FWHM is connected with velocity of gas as:

$$\textrm{FWHM}\approx v_\textrm{obs}=v_\textrm{gas}\cos{\theta},$$
where $v_\textrm{obs}$ is projected velocity on the LOS. The line width depends on contribution of the approaching and receding velocities (shown in Fig. \ref{fig1} as vertical blue and red arrows, respectively), where the clouds closer to the SMBH will contribute to the far line wings, and clouds which are farther from the SMBH contribute to the central part of the line (in Fig. \ref{fig1} upper panel, above the line profile is shown color which corresponds to the clouds color).

Eq. (1) can be used in the case in the case of  fully virialized emitting gas, and also if the projected velocity corresponds to the total velocity, i.e. from the angle of observation. On the other hand, to measure the mass of the SMBH, one should know the distance of the emitting cloud (or emitting clouds, see  Fig \ref{fig1}). In principle in the case of the BLR, there are a group of emitting clouds with geometry that may be disc-like, but also there can be different geometries (see e.g. [24]). Therefore, the SMBH mass is obtained using following relation (see e.g. [8]):

$$M_\textrm{\textrm BH}=f\cdot {v^2_\textrm{obs} R_\textrm{BLR} \over G}, \eqno(2) $$
where $f$ is the virial factor which includes the inclination and geometry of the BLR.

As we noted above the velocity of gas can be estimated from FWHM of a broad line, and second problem is to estimate the dimension of the BLR ($R_\textrm{BLR}$) which emits the broad line.

One of the first ideas was to use the luminosity of broad lines, since the dimension of the BLR is in close connection with the luminosity. Here we simply (from historical reasons) mention, 
so called, Dibai method (see [26-28]) that follows a simple logic that luminosity in the broad line should be proportional to the BLR volume, and consequently 
$$R_\textrm{BLR} \sim L^{1/3}_\textrm{line},$$
and then the SMBH mass can be obtained as (see [28]):
$$M_\textrm{BH}={3\over 2}{R_\textrm{BLR} v^2\over G},$$
where the  velocity of the gas, $v$, was estimated from FWHM (see more details in [26]). Using this, very simple logic, estimated value of the black hole masses were in a relatively good agreement with more precise reverberation method (see discussion in [28]).

More precise method is to determine $R_\textrm{BLR}$ from the variability, so called reverberation method, that was also given as an idea in 1980s (see  [29-32]).
Reverberation  is a very effective way to estimate dimensions of the BLR since the most of AGNs show variability in the spectral lines and continuum. Assuming that the continuum source is very compact, the averaged BLR size corresponds to delay between variability in the continuum and in the broad line flux, i.e. 
$$R_\textrm{BLR}\approx c\Delta\tau,$$
where $c$ is the speed of light and $\Delta\tau$ is the time delay between the continuum and broad line flux variability. This idea has been developed 1990s, from different groups (see e.g. [33-41]), and  the monitoring campaigns continue until today (see e.g. 
[42-53]\footnote{Note here that there are a lot of works showing the results of monitoring campaigns, see e.g. http://www.astro.gsu.edu/AGNmass/}. The reverberation method for BLR dimension  detrmination and consequently  SMBH mass measurements, has been used to find so called radius-luminosity relation (R-L relation,
see e.g. [43,46, 48. 54-60]). The relation in the form of
$$\log(R_\textrm{BLR})=a\cdot\log{L_\textrm{cont}}+b,$$
where $R_\textrm{BLR}$ is  the dimension of the BLR (which is emitting the broad line) and nearby continuum. This connection for different lines is important to measure SMBH masses from an epoch observation for AGNs on different cosmological scales. As e.g. for low-redshifted AGNs, this relation is found for the infrared and optical (mostly H$\alpha$ and H$\beta$) lines. In the UV there are relations for Ly$\alpha$, C IV, C III]  and Mg II lines (see \S 4).

\subsection{Mass of SMBH and line shift}

The gravitational redshift is the second effect that can be produced by the influence of SMBH mass to the broad line profile. If one emitting cloud (EC) is at the distance $R_\textrm{EC}$ from SMBH of mass $M_\textrm{BH}$ (see Fig. \ref{fig1}) it is expected that gravitational redshift, $z_g$  is (see [13,25]):
$$ z_g={GM_\textrm{BH}\over {c^2R_\textrm{EC}}}\eqno(3)
$$
where $c$ is the speed of light.

The gravitational redshift effect mostly can be seen in the case of the broad Fe K$\alpha$ line (see [18-22]), however, one can expect that this effect can contribute to the asymmetry of the UV/Optical lines  (see [13,14,25]). As it is shown in Fig. \ref{fig1}, the clouds which are closer to the central SMBH will have the highest velocities (approaching and receding velocities) and their emission will contribute to the blue/red line wings. If there is the gravitational redshift, it can be seen as a redshift of whole line (see e.g. [25,61,62]), or as a strong asymmetry (as it shown in Fig. \ref{fig1} upper right panel), and there is difference in so called intrinsic redshift of a line, i.e. the redshift difference between the FWHM and Full With at 10\% of Maximal Intensity -- FW10\%M (see e.g. [13] and [14]). Note here that in the case of the gravitational redshift one can measure the asymmetry at different maximal intensity of the line, taking that the center of line (coming from clouds farther from the SMBH, see Fig. \ref{fig1}) has negligible gravitational redshift, and that gravitational redshift affects the line wings (at 5\% and 10\% maximal intensity, FW5\%M and FW10\%M, respectively -- see upper panel  in Fig. \ref{fig1}). 

\subsection{BLR virialization and geometry}

One of the big problem in the SMBH mass measuring using broad lines is the assumption that the BLR is virialized. One can expect that in the strong gravitational field the motion of the clouds is driven by the gravitational force. However, one cannot exclude some other strong motions, as e.g. outflows that mostly is present in the HILs (see e.g. [63] in more details), but also outflows can be present in the LILs (see e.g. [14] and [64]).

In the case of strong outflows, the part of line profile can come from the region that is not virialized. In that case using the broad lines as tool to measure of SMBH mass cannot give some valid results. In principle the systematic redshift is not dominant, and consequently the gravitational redshift cannot be measured. Contrary, in HIL, as e.g. in C IV a blue shift mostly can be observed, i.e. outflowing emitting region in this line is present [65].

As it is seen in Fig. \ref{fig1}, the angle of view is very important for projected velocities in the emitting region, i.e. the inclination of the emitting region is also important for using Eq. 2. Additionally, beside the rotating-like region (assumed to be virialized) there  can be additional emission from a region that is not virialized and this should be taken into account. These effects are assumed to be accounted in the virial factor $f$ given in Eq. 2 (more  about the virial factor $f$ see in \S 4).

One of the methods which can be applied to explore virialization is to compare the widths and intrinsic shifts of broad lines (see e.g. [13] and [14])

\section{Broad spectral lines and SMBH measurements}

In  Table \ref{line-list} we give a list of broad lines (with useful references) which have been used in the SMBH mass measurements. In this part we will shortly discuss validity to use each of the lines from the list for SMBH mass estimates. 

\subsection{Fe K$\alpha$ line}

Iron Fe K$\alpha$ is assumed to be originate in the innermost part of the accretion disk
[18-20] and should reflect the mass of the black hole. There are some effects that can be clearly seen in the Fe  K$\alpha$ line profiles. The specific Fe K$\alpha$ shape indicates an accretion disc line origin [19-22]  that should be close to the SMBH. This implies that the line could be used for SMBH mass measurements [66-67],
however there are several problems to use the Fe K$\alpha$ line for SMBH measurements. First problem is that line is originated in an accretion disc, and the line parameters are strongly depending from the disc parameters (see [19-22] and [66,67]). Second  problem is that is very hard to determine the distance of the Fe K$\alpha$ emitting region, since the reverberation in Fe K$\alpha$ line (see e.g. [68-70]) cannot give distance of the line emitting region. Simply, the line is formed in the region where also X-ray continuum is originating, therefore, we cannot expect to have time delay as it is expected in the optical and UV lines.

Variability in the Fe K$\alpha$ line intensity and shape is connected with different processes in the innermost part of an AGN. In principle Fe K$\alpha$ line can be formed in three distinct regions: (i) in the accretion disc that is very close to the SMBH; (ii) as a reflection from the X-ray corona; and  (iii) in the outflowing region. It means that problem is also with virialization of the line, especially in the part that is emitted in the outflowing region..

Taking all stated above, one can conclude that the Fe K$\alpha$ (especially with resolution that can observed with present day X-ray telescope) is not  good enough for SMBH mass estimates. 

\subsection{The UV lines}

 Using the UV lines one is able to estimate SMBH masses at larger cosmological scales (higher redshifts), since the cosmologically redshifted UV lines can be observed in the optical/infrared spectral bands.  Mostly  Ly$\alpha$, CIV, C III]  (HILs) and Mg II (LIL) lines have been used for SMBH mass estimates. Also, recently a method  that uses the UV Fe III $\lambda\lambda$2039-2113 lines has been proposed (see [61,62]). 
 
\subsubsection{Ly$\alpha$ line}

The Ly$\alpha$ line is a bright line in AGN spectra, and therefore expected to be used in the SMBH mass estimates. The line is in the UV spectrum ($\lambda=1215$\AA ), and can be detected in the case of high redshifted AGNs, i.e. can be used to measure SMBH masses at larger cosmological scales, that is important for investigation of galaxy evolution and cosmology.

There are several reverberation campaigns (see e.g. [71-76]) where a relatively small time lag has been found. This implies that the Ly$\alpha$ emitting region is small (a couple days in the dimensions). The monitoring campaigns provide the relationship between the dimensions of  BLR and the nearby continuum luminosity. Recently, it has been found the relation between BLR dimension and the continuum luminosity at $\lambda$1345\AA\ (see [76]) as:
$$\log(R_{Ly\alpha BLR})=(0.61\pm 0.80)+(0.35\pm0.19)\log(L_\textrm{1345}),$$
where $R_{Ly\alpha BLR}$ is photometric dimension of the Ly$\alpha$ BLR given in 10 light years, 
$L_\textrm{1345}$ is given in 10$^{43}$ erg s$^{-1}$. The BLR dimension in combination with FWHM (see Eq. 2) can be used to SMBH estimates.

We should note here that Ly$\alpha$ line has a high atomic probability, this may cause that a part of line emission can be originated in a non-virialized region. Also,  the Ly$\alpha$ line could be partly absorbed that can affect the measured line FWHM. Also, there is a question about validity of reverberation to determine the sizes of Ly $\alpha$ BLR, since we cannot assume that the continuum source is point like (with respect to the Ly$\alpha$ BLR dimensions). All of these facts are not in favor to use the line for SMBH mass estimation, but in any case using this line we can at least have some indication about SMBH masses of distant AGNs. 

\subsubsection{C IV line}

The C IV line, as a rule shows a blue asymmetry and the line is often shifted to the blue (see [65]), however the line is very bright in the spectra of high-redshifted quasars and it is very often used to estimate their masses.

One of the ways to calibrate the C IV line for mass determination is to use measurements for H$\beta$ and Mg II lines  which are assumed to be more virialized, i.e. the influence of an outflowing region in these lines is not so important as it is in the C IV one. However, comparing the Mg II and C IV in a large sample of quasars, it was found   that there is a weak correlation between the widths of these two lines [77]. It means that C IV is probably weakly virialized. There are several attempts to calibrate the C IV line using Balmer lines (see e.g. [78]), and there are several relations between the C IV line width, nearby continuum (at $\lambda$1350\AA\ or $\lambda$1450\AA) and SMBH masses, which can be used for the SMBH mass estimates using the C IV line (see e.g. [56,78,79]).
It is widely used relation given by Vestergaard \& Peterson [56]:
$$\log(M_\textrm{BH})= (6.66\pm0.01)+ 2\cdot\log(FWHM_\textrm{CIV})+$$
$$+0.53\cdot\log(\lambda L_{\lambda1350\AA}),  $$
 $FWHM_\textrm{CIV}$ is  given in 10$^3$ km s$^{-1}$ and $L_{\lambda1350\AA}$ is the luminosity at $\lambda 1350$\AA\ given in 10$^{44}$ erg s$^{-1}$ 

There are several works which try to improve the relation between the C IV width (or velocity dispersion), the local continuum  and black hole masses (see [79-82]). The main problem is that there is contribution of non-virialized emission, i.e.  a blue shift in C IV broad line is (as a rule) present
(see e.g [77,83]), which is indicating an outflowing region.
Therefore,  there are several attempts to find a correction that the C IV can be fully used for SMBH mass estimates (see 78,79,82,84].

As an example,  Mej\'ia et al. [84] gave some improvements of the relation using the luminosity ratios. The relation for SMBH mass estimate (using C IV line parameters) that was calibrated to the Mg II mass measurements taking C III] and C IV line ratio is given  as [83]:

$$M_\textrm{BH}=5.71\cdot 10^5\cdot L^{0.57}_\textrm{1450}\cdot FWHM(CIV)^2\cdot$$
$$\cdot\Big({L_P(CIII])\over {L_P(CIV)}}\Big)^{-2.09}, $$
where $L_\textrm{1450}$ is the continuum luminosity at $\lambda$1450\AA\ given in 10$^{44}$ erg s$^{-1}$, $ FWHM(CIV)$ is the FWHM of C IV line given in 10$^3$ km s$^{-1}$, $L_P(CIII])$ and $L_P(CIV)$ are the intensity of C III]  and C IV lines, respectively.

This and similar approaches (see e.g. [77,85-87]) which include additional spectral characteristics where inspired by the Baldwin effect [88], where C IV line shows an anti-correlation with the local continuum, i.e. show some physical  connections between the different spectral characteristics.

Note here that for C IV BLR dimension estimates. the continuum luminosity at $\lambda$1345\AA\ have been used (see [76]) and  also the luminosity of C IV line [89]. As e.g. in [89] the relation between C IV BLR dimensions and line luminosity is given as:
$$\log\Big(R(BLR)_\textrm{CIV}\Big)= 0.69 + 0.6 \log\Big({L_\textrm{CIV}\over{ 10^{42}}}\Big),$$
where $R(BLR)_\textrm{CIV}$ is given in light days. The above equation with the FWHM of C IV can be used (in Eq. 2) to estimate SMBH masses.

Taking the discussion above we can say that the C IV line is not perfect for the SMBH mass measurements, but still can be used with some modifications, where the line peak intensity of different lines have been included (see also [86]).

\subsubsection{CIII] line}

The  CIII]$\lambda1909$\AA\ line in spectra of AGNs is weaker than C IV, but still it has been considered to be used  for SMBH mass estimates (see [76,84,90]). The problem with the line is that in its blue wing there are two satellite lines AlIII]$\lambda1857$\AA\ and Si III]$\lambda1892$\AA. There are several investigation in order to find the dimensions of these regions and to use the line for determination of the SMBH masses.
It seems that the emission regions are several light days in dimensions (see e.g. [89]). As e.g. the dimension of the C III] BLR ($R_\textrm{C III]}$ in light days) as a function of the luminosity at $\lambda$1350\AA\ (L$\lambda$1350\AA\ in 10$^{43}$  erg s$^{-1}$)   is given in [76] as:
$$R_\textrm{C III]}=(1.10\pm0.77)\cdot(\lambda L_{\lambda1350\AA })^{(0.26\pm0.16)}.$$
This estimate of the BLR dimensions can be used (see Eq. (2)) to estimate SMBH masses. However, there is a large uncertainty coefficients in equation above, and consequently in the $R_\textrm{C III]}$ determination. Taking this into account and that the C III] line is contaminated with Al III] and Si III] lines (i.e. that can affect FWHM measurements), the line is not often used for SMBH mass estimates.

\subsubsection{Fe III $\lambda\lambda$2039-2113\AA\ line blend}

The line shift is not often used to measure the masses of SMBHs, but we can expect that an emitting region located close to the SMBH can have a gravitational redshift (see [13,25,91]). This, recently, has been used to interpret the shift of the Fe III $\lambda\lambda$2039-2113\AA\ line blend as a gravitational redshift by Mediavilla et al. [61,62]. They found a relationship between the shift of the Fe III bland ($z_\textrm{FeIII}$), the continuum luminosity at $\lambda$1350 and black hole mass as (62]:
$$M_\textrm{BH}=10^a z_\textrm{FeIII} \cdot (\lambda L_{\lambda1350\AA })^b,$$
where the SMBH mass is given in Solar masses, $z_\textrm{FeIII}$ in 10$^3$ km s$^{-1}$, $L_{\lambda1350\AA }$  in 10$^{41}$ erg s$^{-1}$ and coefficients are: 
$a=7.89^{-0.11}_\textrm{-0.13}$ and $b=(0.57\pm 0.08)$.

The method has been tested and seems to give a reasonable accuracy with respect to other methods (see in more details [62]).

\subsubsection{Mg II line}

There are a number of papers discussing the Mg II $\lambda2800$\AA\ line as the SMBH mass estimator (see e.g. [13,14,56,62,76,84,92-97]. It seems that the Mg II line is one of the best UV lines for SMBH mass measurements, and it has been used for estimates of the SMBH masses of higher cosmological redshifted quasars. But also the Mg II line has been used for the calibration other UV lines (first of all C IV line).

The relations between the SMBH mass, luminosity at $\lambda$3000\AA\ ($L_{\lambda3000\AA}$) and Mg II FWHM were given in several papers (starting from relation given in [56] to one given in [97]). 
In principle the SMBH masses from Mg II line seem to be consistent with virial assumption as (see [96,97]):
$$M_\textrm{BH} \sim (L_{\lambda3000\AA})^{0.5}\cdot FWHM_\textrm{MgII}^2,$$
that is often presented in the the log-scale as:

$$\log(M_\textrm{BH})= \alpha+\beta\cdot\log(L_{\lambda3000\AA})+\gamma\cdot \log(FWHM_\textrm{MgII}),$$
  where $\alpha,\ \beta$ and $\gamma$ are constants which are derived from comparison of the H$\beta$ line and Mg II line in a number of AGNs (see e.g. [92-97]). As we mentioned there are different estimates for constants, as an example in [93] by fitting a number of AGNs they give following constant values: 
  $\alpha=1.15 \pm 0.27$,  $\beta=0.46 \pm 0.08$, and  $\gamma =1.48 \pm 0.49$, where the mass is given in 10$^6 M\odot$, the  $FWHM_\textrm{MgII}$  is given in  10$^3$ km s$^{-1}$, and the  $L_{\lambda3000\AA }$  in 10$^{44}$ erg s$^{-1}$.
  
It seems that the FWHM of Mg II is systematically smaller than H$\beta$ line, but  still can be calibrated using the H$\beta$ and give good results for SMBH masses. However, one should be careful with using Mg II line, since the line profiles seem to be different than H$\beta$ region (especially in the line wings, see e.g. 13,14]) and in some cases the line variability indicates that the line is not originated in the viralized region (see e.g. [61,62]). Also, there can be problem with very broad Mg II lines (FWHM$_\textrm{MgII}>$6000 km s$^{-1}$, see [98]) where a non-virialized emission can be dominant (see [14]). Additionally, from the reverberation of the Mg II line and local continuum (at $\lambda3000\AA$) it is clearly seen that response the Mg II to the continuum variability is smaller than in the case of 
Balmer lines (see [99-102]). All of these facts indicate that one should be very careful when using the Mg II line for SMBH mass estimates.

\subsection{Optical lines}

\subsubsection{H$\beta$ and H$\alpha$ lines}

In the optical part of AGN spectra, the Balmer lines are dominant, first of all, broad H$\alpha$ and H$\beta$ lines.
It seems that H$\alpha$ and H$\beta$ broad lines are virialized, and that they are originated in a flattened BLR (disk-like that is close to the face-on orientation to the observer, see e.g [103]). However, the reverberation mapping of the H$\beta$ and H$\alpha$ emitting region shows that they are probably different in dimensions (see e.g. 
[104-107]). Note here that H$\alpha$ line has a higher atomic probability than H$\beta$ and there is always possibility that some portion of H$\alpha$ emission is coming from a non-virialized region, therefore, the H$\beta$ seems to be more reliable for the SMBH mass measurements than H$\alpha$.

The main goal of reverberation mapping of Balmer lines, mostly the broad H$\beta$ line, is to clarify the BLR geometry and dimensions in order to use Eq. 2 for mass determination. First of all the time delay can be obtained from the cross correlations between the continuum and broad line variability in order to find the BLR dimensions, i.e. $R_\textrm{BLR}-L_{\lambda5100\AA}$ relation. This relation (usually find for H$\beta$) have been using for single epoch SMBH measurements and for calibration of the UV lines (first of all C IV and Mg II, see text above).

The  $R_\textrm{BLR}-L_{\lambda5100\AA}$ relation can be written as [108]:

$$\log(R_\textrm{BLR})= A+\alpha\cdot\log(\lambda L_{\lambda5100\AA}),$$
where $A$ and $\alpha$ are constants, which are mostly obtained from fitting the obtained time lags. As e.g.  Bentz  et al. [109] gave: $A= -21.3^{+2.9}_{-2.8}$ and 
$\alpha= 0.519^{+0.063}_{- 0.066}$.

Second problem to use Eq. 2 is to find  virial factor $f$, that depends on the BLR geometry and inclination. There are several estimates for factor $f$ that has values between 2.8 to 5.5 (see [12,110-113], and critical discussion in [12,114]).

The Balmer lines seem to be very good tool for SMBH mass determination, however, there are also some problems with so called narrow line Sy 1 (NLSy1, see e.g. [115,116]) or high accreting AGNs.

\subsection{Infrared lines: Pa$\alpha$, P$\beta$, Br$\alpha$ and Br$\beta$ lines}

The Pashen and Brecket line series can be observed in the near infrared AGN spectra and have been  used for determination of the SMBH masses (see [117-121]).  The infrared lines are interesting for SMBH mass determination since we expect that  50\% AGNs are red and obscured with dust (see [117]). Such obscured AGNs still can show the broad line emission in the infrared part of spectra, i.e. they show typical spectra for Type II objects (narrow lines) in the optical part. The masses of these galaxies can be determined using the infrared spectra.

Prominent lines in the infrared part of AGN spectrum are Pa$\alpha$,  Pa$\beta$, Br$\alpha$ and Br$\beta$ lines, and the calibration of the relations between nearby continuum (or line) luminosity, line width (or velocity dispersion) and SMBH mass is based on the relations obtained for the optical broad lines, first of all by using the H$\beta$ broad line.

The relation between the SMBH mass and the continuum luminosity at 1$\mu$m and width of Pa $\alpha$ is given as (see [118]:
$$\log(M_\textrm{BH}) = a\cdot(2 \log(FWHM) + 0.5 \log(\nu L_{1\mu m})-b,$$
where $a=0.88 \pm 0.04$ and $b=(17.39 \pm 1.02) $, and $M_\textrm{BH}$ is given in solar masses.

There are relations between the SMBH mass (in solar masses), the luminosities of  Pa$\alpha$ and
Pa$\beta$ and their widths (see [117]) that can be written as 
$$\log(M_\textrm{BH}) = a+b\cdot\log(L_\textrm{Pan})+ c\cdot\log(FWHM_\textrm{Pan})\ \ (n=\alpha,\beta),$$
where FWHM$_\textrm{Pn}$ is given in 10$^3$ km s$^{-1}$ and $L_\textrm{Pan}$ is given in 10$^{42}$ erg s$^{-1}$. It is expected that $b=0.5$ and $c=2$ (see [26,27]). The values  for the coefficients are given in [117], and they have values for Pa$\alpha$: $a=7.29,\ b=0.43$ and $c=1.92$, while for Pa$\beta$: $a=7.33,\ b=0.45$ and $c=1.69$. As it can be seen the values are not so far from those which are expected in the case of photoionization.

In the case of Br$\alpha$ and Br$\beta$, the relation between line luminosity and dispersion in the line is given as (see [119]):
$$\log(M_\textrm{BH}) = a_n\cdot(2 \log(\sigma_\textrm{line}) + b_n \log(\nu L_\textrm{n})\ \ \ (n=\alpha,\beta),$$
where $\sigma_n$ is given in 10$^3$ km s$^{-1}$, and $L_n$ is given in 10$^{40}$ erg s$^{-1}$ and $M_\textrm{BH}$ is given in solar masses. The coefficients for Br$\alpha$ are $a_\alpha=(6.68\pm 0.20)$ and $b_\alpha=(0.66\pm 0.21)$, while for Br$\beta$ are $a_\beta==(6.61\pm 0.23)$ and $b_\beta=(0.68\pm0.13)$.

On the other hand, the measurements of SMBH masses in Type 2 objects also can be combined the widths of infrared lines  (Pa$\alpha$, Pa$\alpha$, and HeI $\lambda$1.083 $\mu$m), since they have similar FWHM in combination with the X-ray emission (see e.g.
[120,121]). The X-ray luminosity  L$_\textrm{14-195\ keV}$ in the energy interval from 14 keV to 195 keV in combination with FWHM$_\textrm{IR}$ of the infrared lines can give the mass of black hole as (see [121]):
$$\log(M_\textrm{BH})=7.75+2\cdot\log(FWHM_\textrm{IR}+0.5\cdot\log(L_\textrm{14-195keV}),
$$
where $M_\textrm{BH}$ is given in solar masses, FWHM$_\textrm{IR}$ in 10$^4$ km s$^{-1}$ and L$_\textrm{14-195keV}$ in the 10$^{42}$ erg s$^{-1}$.

Note here that in order to measure the SMBH masses of Type 2 AGNs, some connections (correlations, ratios) between the narrow and broad line are explored, then using only the narrow lines (observed in Type 2 AGNs) one can obtained their SMBH masses. As an example mention here recent work given in [122], where a correlation between the luminosity of narrow [OIII] and H$\beta$ lines (L([O III])/L(H$\beta$) line ratio and the FWHM H$\alpha$ has been used to find SMBH masses in Type 2 AGNs.

\subsection{Polarization in broad lines - a method for SMBH mass measurements}

It is expected that polarization in broad lines is caused by the equatorial scattering in the inner part of the torus (see [123]). In the case of Keplerian-like motion, it is expected that the shape of polarization angle is depending from the velocities across the line profile (see [123,124]). It was shown in [15] that the Keplerian-like motion in 
combination with equatorial scattering give a velocity distribution across the line profile as:

$$\log({V_i\over c})=a-b\cdot \log(\tan(\Delta\varphi_i)), \eqno(5)$$
where $c$ is the speed of light, $\Delta\varphi_i$ is the value of polarization angle across the line profile. The constant $a$   depends on the SMBH mass ($M_\textrm{BH}$) as
$$a=0.5\log\bigl({{GM_\textrm{BH} \cos^2(\theta)}\over{c^2R_\textrm{sc}}}\bigr). \eqno(6)$$
where $G$ is the gravitational constant, $R_\textrm{sc}$ is the inner part of the tours where is the scattering region located. $\theta$ is the  inclination angle between the torus and BLR. In principle it can be considered that the BLR and torus are co-planar, and consequently  $\theta\sim0$. In the case Keplerian-like motion  $b=0.5$.

The equation above has been tested in a group of broad line AGNs (see [15,17]), where the measurements of SMBH masses with polarization method were compared with other methods (reverberation and $M_\textrm{BH}-\sigma$ method), and it was found that the polarization method gives very good SMBH mass estimates. On the other hand, the limitation of the method was investigated by Savi\'c et al. [16] using the STOKES code, and it was found that the method can be used in the case where, beside Keplerian-like motion, also some outflows/inflows exist in the BLR. The method is robust and can be applied for one epoch observations, and also can be applied for different emission lines. In the case that Keplerian-like motion is not present, the method cannot be used. However, the presence of Keplerian-like motion can be simply detected in the polarization angle shape across the line profile. This is an advantage of the method, that there is no assumption of virialization, as it is mostly the case in the reverberation method.

Note here, that there are some works trying to estimate virial factor $f$ using polarization (see [125,126]).

\section{Conclusion}

Here we present methods for measurements SMBH masses in AGN using broad emission lines.
We discuss the lines from X-ray to the infrared spectra, and give  an overview  of the broad lines and their possibility for SMBH mass measurements. Additionally we give some of the useful relations between the SMBH mass and spectral line parameters.

To summarize discussion above we point out following:

\begin{itemize}

\item{} The Fe K$\alpha$ line seems to be very good for the investigation of accretion disc structure, and potentially can be used for spin determination, however, due to low resolution of X-ray telescopes, and the fact that line is originating very close to the central SMBH, it is very hard to use this line for SMBH mass determination.

\item{} The UV lines can be used for the SMBH mass determination, however, Ly$\alpha$, C IV and C III] lines are affected by some absorption and outflows (especially C IV) that could be used with caution. The best seems to be Mg II line, but also there are some problems that one has to be careful to use this line for SMBH mass measurements.

\item{} The optical H$\alpha$ and H$\beta$ lines seem to be very promising for SMBH mass measurements, especially H$\beta $ line, and there are several estimates of the virial factor and $R-L$ relations that can be used for SMBH measurements. Also, in this case, one should be careful, since there may be a problem with different sub-types of broad line AGNs, as well high accreting AGNs seems to have different $R-L$ relations.

\item{} The IR lines are very useful for SMBH mass measurements, especially in the Type 2 AGNs, where still Pashen and Brecket broad lines are presented. The IR lines give a unique opportunity to measure SMBH mass in Type 2 AGNs.

\item{} Polarization in broad lines seems to be a promising method for measurements of SMBH masse, however, to use the method the high performed polarization observation are required.
\end{itemize}

At the end, let us note that with new generation of telescopes we expect to be able to map the motion of the gas very close to the black hole, i.e. clearly measure gradient of velocities in clouds around an SMBH (see Fig. \ref{fig1}). Recently it was performed by 
Gravity Collaboration (see [127]), where Pa$\alpha$ from the very central part of 3C 273
was observed (with a spatial resolution of $\sim$36 light days), this allows to measure the gas velocity very close to the SMBH, and they found that the mass of this AGN is $3\cdot10^8M\odot$. Such precise observations will be used to calibrate UV lines in order to find SMBH masses of distant (high redshifted) quasars.

\section{Acknowledgement}

This work is a part of the project (176001) "Astrophysical Spectroscopy of Extragalactic Objects," supported by the Ministry of Science and Technological Development of Serbia. 
The review was presented at 12th Serbian Conference on Spectral Line Shapes in Astrophysics in the special session "Broad lines in AGNs: The physics of emission gas in the vicinity of super-massive black hole" (In memory of the life and work of dr Alla Ivanovna Shapovalova). 

\section{References}
\begin{itemize}

 \item[1.] The Event Horizon Telescope Collaboration, 2019, The Astrophysical Journal Letters, 875:L1 (17pp)
 \item[2.]  Ferrarese, L. Merritt, D. 2000, ApJ, 539L, 9-12
\item[3.] Kormendy, J., Ho, L. C. 2013, ARA\&A, 51, 511-653
\item[4.] Bentz, M. C., Manne-Nicholas, E. 2018, ApJ, 864, article id. 146, 19 pp.
\item[5.] Davis, B. L., Graham, A. W., Cameron, E. 2019, ApJ, 873, article id. 85, 26 pp.
\item[6.] Alexander, T. 2017, ARA\&A, 55, 17-57
\item[7.] Naab, T. O., Jeremiah P. 2017, ARA\&A, 55, 59-109 
\item[8.] Peterson, B. M.  2014, SSRv, 183, 253-275
\item[9.] Tremaine, S., Gebhardt, K., Bender, R., Bower, G. et al. 2002, ApJ, 574, 740.
\item[10.] G\"ultekin, K., Richstone, D. O., Gebhardt, K. et al. 2009, ApJ 698,  198
\item[11.] Graham, A. W., Onken, C. A., Athanassoula, E., Combes, F. 2011, MNRAS, 412, 2211.
\item[12.] Shankar, F., Bernardi, M., Richardson, K. et al. 2019, MNRAS, 485, 1278
\item[13.] Joni\'c, S., Kova\v cevi\'c-Doj\v cinovi\'c, J., Ili\'c, D., Popovi\'c, L. \v C. 2016, Ap\&Sp. Sci, 361,  id.101, 24 pp.
\item[14.] Popovi\'c, L. \v C., Kova\v cevi\'c-Doj\v cinovi\'c, J., Mar\v ceta-Mandi\'c, S. 2019, MNRAS, 484, 3180
\item[15] Afanasiev, V. L., Popovi\'c, L. \v C. 2015, MNRAS, 800,  id.L35, 4 pp. 
\item[16.] 	Savi\'c, D., Goosmann, R., Popovi\'c, L. \v C., Marin, F., Afanasiev, V. L. 2018, A\&A, 614, id.A120, 16 pp.
\item[17.] Afanasiev, V. L., Popović, L. \v C., Shapovalova, A. I. 2019, MNRAS, 482, 4985 
\item[18.] Tanaka, Y., Nandra, K., Fabian, A. C. et al. 1995, Nature,  375, 659 
\item[19.] Fabian, A. C., Nandra, K., Reynolds, C. S., Brandt, W. N., Otani, C., Tanaka, Y., Inoue, H., Iwasawa, K. 1995, MNRAS, 277,  L11.
\item[20.] Nandra, K., George, I. M., Mushotzky, R. F., Turner, T. J., Yaqoob, T. 1997, ApJ, 477,  602
\item[21.] Jovanovi\'c, P. 2012, New AR,  56,  37 
\item[22.] Milo\v sevi\'c, M., Pursiainen, M. A., Jovanovi\'c, P., Popovi\'c, L. \v C. 2018, International JMPys. A,  33,  id. 1845016
\item[23.] Jiang, J., Walton, D. J., Fabian, A. C., Parker, M. L. 2019, MNRAS, 483, 2958
\item[24.]	Sulentic, J. W., Marziani, P., Dultzin-Hacyan, D. 2000, ARA\&A,  38, 521 
\item[25.] Popovi\'c, L. \v C., Vince, I., Atanackovi\'c-Vukmanovi\'c, O., Kubi\v cela, A. 1995, A\&A, 293, 309
\item[26.] Dibai, E. A. 1977, SvAL, 3, 1-3
\item[27.] Dibai, E. A. 1978, SvA, 22, 261-266
\item[28.] Bochkarev, N. G.,  Gaskell, C. M. 2009, Ast. Lett., 35, 287-293 
\item[29] Blandford, R. D. \& McKee, C. F. 1982, ApJ, 255, 419
\item[30]  Capriotti, E. R., Foltz, C. B., Peterson, B. M. 1982, ApJ, 261, 35
\item[31]  Gaskell, C. M. \& Sparke, L. S. 1986, ApJ, 305, 175
\item[32]  Wandel, A.\& Yahil, A. 1985, ApJ, 295L, 1
\item[33] Clavel, J., Reichert, G. A., Alloin, D. et al. 1991, ApJ, 366, 64
\item[34] Peterson, B. M., Balonek, T. J., Barker, E. S. et al. 1991, ApJ, 368, 119 
\item[35] Peterson, B. M., Alloin, D., Axon, D. et al. 1992, ApJ, 392, 470
\item[36] Dietrich, M.,  Kollatschny, W.,  Peterson, B. M. et al. 1993, ApJ, 408, 416
\item[37] Stirpe, G. M., Winge, C., Altieri, B et al. 1994, ApJ, 425, 609
\item[38] Rodríguez-Pascual, P. M., Alloin, D., Clavel, J. et al. 1997, ApJS, 110, 9
\item[39] Collier, S. J., Horne, Keith,  Kaspi, S.  et al. 1998, ApJ, 500, 162
\item[40] Doroshenko, V. T., Sergeev, S. G., Pronik, V. I., Chuvaev, K. K. 1999, AstL, 25, 569
\item[41]  Wandel, A., Peterson, B. M., Malkan, M. A.  1999, ApJ, 526, 579
\item[42] Shapovalova, A., Burenkov, A., Spiridonova, O. et al. 2000, ASPC, 215, 219
\item[43] Kaspi, S., Smith, P. S., Netzer, H. Maoz, D., Jannuzi, B. T., Giveon, U. 2000, ApJ, 533, 631
\item[44] Shapovalova, A. I., Burenkov, A. N., Carrasco, L. et al. 2001, A\&A, 376, 775 
\item[45] Tao, J., Qian, B., Fan, J. 2004, PASP, 116, 634
\item[46] Bentz, M. C., Denney, K. D., Cackett, E. M. et al. 2006, ApJ, 651, 775
\item[47] Denney, K. D.,  Watson, L. C., Peterson, B. M. et al. 2009, ApJ, 702, 1353
\item[48] Bentz, Misty C., Walsh, J. L., Barth, A. J. et al. 2009, ApJ, 705, 199
\item[49] Popovi\'c, L. \v C., Shapovalova, A. I., Ili\'c, D. et al. 2014, A\&A, 572A, 66
\item[50] Du, P., Hu, C. Lu, K.-X. et al. 2015, ApJ, 806, 22 
\item[51] Shapovalova, A. I., Popovi\', L. \v C., Chavushyan, V. H.  et al. 2016, ApJS, 222, 25 
\item[52] Shapovalova, A. I., Popovi\'c,  L. \v C., Afanasiev, V. L. et al, 2019, MNRAS,485, 4790 
 \item[53] Zhang, Z.-X., Du, P., Smith, P. S. et al. 2019, ApJ, 876, 49
 \item[54] Koratkar, A. P., Gaskell, C. M. 1991, ApJ, 370L, 61 
\item[55] Kaspi, S., Maoz, D., Netzer, H. et al. 2005, ApJ, 629, 61 
\item[56] Vestergaard, M., Peterson, B. M. 2006, ApJ, 641, 689 
\item[57] Landt, H., Bentz, M. C., Peterson, B. M. et al. 2011, MNRAS, 413L, 106
\item[58] Bentz, M. C., Denney, K. D., Grier, C. J. et al. 2013, ApJ, 767, 149 
\item[59] Kilerci E., E., Vestergaard, M., Peterson, B. M., Denney, K. D., Bentz, M. C.2015, ApJ, 801, 8 
\item[60] Du, P., Lu, K.-X., Zhang, Z.-X. et al. 2016, ApJ, 825, 126 
\item[61] Mediavilla, E., Jm\'enez-Vicente, J., Fian, C.  et al. 2018,  2018, ApJ,862, 104 
 \item[62] Mediavilla, E., Jimn\'ez-Vicente, J., Me\'ja-Restrepo, J., Motta, V., Falco, E., Mu\~noz, J. A., Fian, C., Guerras, E. 2019,  ApJ, 880, 96
\item[63] Marziani, P., del Olmo, A., Martínez-Carballo, M. A. et al. 2019, A\&A, 627A, 88
\item[64] Le\'on-Tavares, J., Chavushyan, V., Pati\~no-Álvarez, V. et al. 2013, ApJ, 763L, 36 

 \item[65] Richards, G. T., Vanden Berk, D. E., Reichard, T. A. et al. 2002, AJ, 124, 1 
\item[66] 	Matt, G., Perola, G. C. 1992, Monthly Notices of the Royal Astronomical Society (ISSN 0035-8711), vol. 259, no. 3, p. 433-436. 
\item[67] Dov\v ciak, M., Bianchi, S., Guainazzi, M., Karas, V., Matt, G. 2004, MNRAS.350, 745
\item[68] Campana, S., Stella, L. 1995, MNRAS, 272, 585
\item[69] Zoghbi, A., Fabian, A. C., Reynolds, C. S.,  Cackett, E. M.	2012, MNRAS, 422, 129
\item[70] Taylor, C., Reynolds, C. S. 2018, The Astrophysical Journal, Volume 868, Issue 2, aid. 109, 17 pp.
\item[71]  Ulrich, M. -H., Courvoisier, T. J. -L., Wamsteker, W. 1993, ApJ, 411, 125 
\item[72] Reichert, G. A., Rodriguez-Pascual, P. M., Alloin, D. et al. 1994,ApJ, 425, 582 
\item[73] O'Brien, P. T., Dietrich, M., Leighly, K. et al. 1998, ApJ, 509, 163 
\item[74] De Rosa, G., Peterson, B. M., Ely, J. et al.  2015, ApJ, 806, 128
\item[75] Goad, M. R.,  Korista, K. T., De Rosa, G. et al. 2016, ApJ, 824, 11
\item[76] Lira, P.,  Kaspi, S., Netzer, H. et al.  2018, ApJ, 865, 56
\item[77] Fine, S., Croom, S. M., Bland-Hawthorn, J., Pimbblet, K. A., Ross, N. P., Schneider, D. P., Shanks, T. 2010, MNRAS, 409, 591 
\item[78] Assef, R. J., Denney, K. D., Kochanek, C. S. et al. 2011, ApJ, 742, 93
\item[79] Mej\'ia-Restrepo, J. E., Trakhtenbrot, B., Lira, P., Netzer, H.  2018, MNRAS, 478, 1929
\item[80.] Park, D., Barth, A. J., Woo, J.-H et al. 2017, ApJ, 839, 93 
\item[81.] Sun, M., Xue, Y.,  Richards, G. T., Trump, J. R., Shen, Y., Brandt, W. N., Schneider, D. P.  2018, ApJ, 854, 128 
\item[82.] Hoormann, J. K., Martini, P., Davis, T. M. et al.2019, MNRAS, 487, 3650 
\item[83.] Ge, X., Zhao, B.-X., Bian, W.-H., Frederick, G. R. 2019, AJ, 157, 148 
\item[84.] Mej\'ia-Restrepo, J. E., Trakhtenbrot, B., Lira, P., Netzer, H., Capellupo, D. M.  2016, MNRAS, 460, 187 
\item[85.] Denney, K. D., 2012, ApJ, 759, 44 
\item[86.] Denney, K. D., Pogge, R. W., Assef, R. J., Kochanek, C. S., Peterson, B. M., Vestergaard, M. 2013, ApJ, 775, 60 
\item[87.] Runnoe, J. C., Brotherton, M. S., Shang, Z., DiPompeo, M. A. 2013, MNRAS, 434, 848 
\item[88.]  Baldwin, J. A. 1977, ApJ, 214, 679
\item[89.]  Kong, M.-Z., Wu, X.-B., Wang, R., Han, J.-L. 2006, ChJAA, 6, 396 
\item[90.] Trevese, D., Perna, M., Vagnetti, F., Saturni, F. G., Dadina, M.  2014,  ApJ, 795, 164 
\item[91.] Liu, H. T., Feng, H. C., Bai, J. M.	 2017. MNRAS, 466, 3323	
\item[92]  Onken, C. A., Kollmeier, J. A. 2008. ApJ, 689L, 13O
\item[93] Wang J.-G., Dong X.-B., Wang T.-G., Ho L. C., Yuan W., Wang H., Zhang K., Zhang S., Zhou H.  2009, ApJ, 707, 1334
\item[94] Marziani, P., Sulentic, J. W., Plauchu-Frayn, I., del Olmo, A., 2013, A\&A, 555A, 89
 \item[95] Kova\v cevi\'c-Doj\v cinovi\'c, J., Mar\v ceta-Mandi\'c, S. Popovi\'c, L.Č.  2017, FrASS, 4, 7
\item[96] Woo, J.-H., Le, H. A. N., Karouzos, M. et al.  2018, ApJ, 859, 138
\item[97] Bahk, H., Woo, J.-H., Park, D.   2019, ApJ, 875, 50
\item[98] Trakhtenbrot, B., Netzer, H. 2012, MNRAS, 427, 3081
\item[99] Cackett, E. M. G\"ultekin, K., Bentz, M. C., Fausnaugh, M. M., Peterson, B. M., Troyer, J., Vestergaard, M.  2015, ApJ, 810, 86
\item[100] Shen, Y., Horne, K., Grier, C. J. et al. 2016, ApJ, 818, 30S
\item[101] Czerny, B.,  Olejak, A., Ra\'cowski, M. et al.  2019, ApJ, 880, 46
\item[102] Yang, Q., Shen, Y., Chen, Y.-C. et al.  2019, arXiv190410912
\item[103] Grier, C. J., Pancoast, A., Barth, A. J., Fausnaugh, M. M., Brewer, B. J., Treu, T., Peterson, B. M.  2017, ApJ, 849, 146
\item[104] Shapovalova, A. I., Popović, L. \v C., Collin, S. et al.  2008, A\&A, 486, 99S
\item[105] Shapovalova, A. I., Popovi\'c, L. \v C., Burenkov, A. N.  et al.  2010, A\&A, 517A, 42
\item[106] Shapovalova, Alla I., Popovi\'c, L. \v C., Chavushyan, V. H. et al.  2017, MNRAS, 466, 4759
\item[107] Jiang, L., Shen, Y., McGreer, I. D., Fan, X., Morganson, E., Windhorst, R. A.  2016, ApJ, 818, 137
\item[108]  Bentz, M. C., Peterson, B. M., Pogge, R. W., Vestergaard, M.,  Onken, C. A.
2006, ApJ, 644, 133
 \item[109] Bentz, M. C., Peterson, B. M.,Netze, H.,  Pogge, R. W.,  Vestergaard, M. 2009, 697, 160  
 \item[110] Graham, A. W., Onken, C. A., Athanassoula, E. Combes, F. 
	2011, MNRAS, 412, 2211 
 \item[111] Grier, C. J.,  Martini, P., Watson, L. C.. et al. 	2013, ApJ, 773, 90
 \item[112] Park, D., Kelly, B. C., Woo, J.-H., Treu, T. 	2012, ApJS, 203, 6
 \item[113] Onken, C. A., Ferrarese, L., Merritt, D. et al. 2004, ApJ, 615, 645
 \item[114] Bentz, M. C., Katz, S. 2015, PASP, 127, 67
  \item[115] Du, P., Zhang, Z.-X., Wang, K. et al. 2018, ApJ, 856, 6 
 \item[116]  Huang, Y.-K., Hu, C., Zhao, Y. et al.  2019, ApJ, 876, 102
\item[117] Kim, D., Im, M., Kim, M.  2010, ApJ, 724, 386
\item[118] Landt, H., Ward, M. J., Peterson, B. M., Bentz, M. C., Elvis, M., Korista, K. T., Karovska, M.  2013, MNRAS, 432, 113 
\item[119] 	Kim, D., Im, M., Kim, J. H. et al. 2015, ApJS, 216, 17
\item[120] Onori, F., Ricci, F., La Franca, F. et al. 2017, MNRAS, 468L, 97
 \item[121] Ricci, F., La Franca, F., Onori, F., Bianchi, S.
    2017, A\&A, 598A, 51
  \item[122]    Baron, D., Ménard, B. 2019, MNRAS, 487, 3404

  \item[123] Smith, J. E., Robinson, A., Young, S., Axon, D. J., Corbett, E. A. 2005, MNRAS, 359, 846 
   \item[124] Afanasiev, V. L., Popovi\'c, L. \v C., Shapovalova, A. I., Borisov, N. V., Ili\'c, D.  2014, MNRAS, 440, 519
 \item[125] Songsheng, Y.-Y., Wang, J.-M.   2018, MNRAS, 473L, 1
\item[126]  Piotrovich, M. Yu., Gnedin, Yu. N., Silant'ev, N. A., Natsvlishvili, T. M., Buliga, S. D.   2015, MNRAS, 454, 1157
\item[127]  Gravity Collaboration: Sturm, E., Dexter, J., Pfuhl, O. et al.  2018, Nature, 563, 657
\end{itemize}

\end{document}